\documentclass[a4paper,twocolumn,accepted=2025-02-10]{quantumarticle}
\pdfoutput=1
\usepackage[utf8]{inputenc}
\usepackage[T1]{fontenc}
\usepackage[english]{babel}
\usepackage{amsmath,amssymb,amsthm}
\usepackage{hyperref}
\usepackage{url}
\usepackage{subcaption}
\usepackage{mathtools}
\usepackage{physics}
\usepackage{booktabs}
\usepackage{datetime}
\hypersetup{
    colorlinks=true,
    linkcolor=blue,
    citecolor=blue,
    filecolor=blue,
    urlcolor=blue
}

\usepackage{verbatim}
\usepackage[T1]{fontenc}
\usepackage{float}
\usepackage{graphicx}
\usepackage{xcolor}
\usepackage{physics}
\usepackage{soul}

\usepackage{bbold}

\definecolor{ao(english)}{rgb}{0.0, 0.5, 0.0}

\theoremstyle{definition}

\newcommand{\bracket}[3]{\langle#1|#2|#3\rangle}

\newcommand{\eps}{\varepsilon}
\usepackage{amssymb}
\usepackage{amsthm}
\usepackage{mathtools} 
\usepackage[customcolors]{hf-tikz} 
\usetikzlibrary{patterns}
\usetikzlibrary{matrix,decorations.pathreplacing}

\pgfkeys{tikz/mymatrixenv/.style={decoration={brace},every left delimiter/.style={xshift=8pt},every right delimiter/.style={xshift=-8pt}}}
\pgfkeys{tikz/mymatrix/.style={matrix of math nodes,nodes in empty cells,left delimiter={[},right delimiter={]},inner sep=1pt,outer sep=1.5pt,column sep=2pt,row sep=2pt,nodes={minimum width=20pt,minimum height=10pt,anchor=center,inner sep=0pt,outer sep=0pt}}}
\pgfkeys{tikz/mymatrixbrace/.style={decorate,thick}}

\usepackage[numbers,sort&compress]{natbib} 

\begin{document}

\definecolor{quantumviolet}{HTML}{555555} 
\definecolor{black}{rgb}{0,0,0} 

	\title{Information capacity of quantum communication under natural physical assumptions}

	\author{Jef Pauwels}
	\affiliation{Department of Applied Physics, University of Geneva, 1211 Geneva, Switzerland}
	\affiliation{Constructor Institute of Technology (CIT), Geneva, Switzerland}
	\author{Stefano Pironio}
	\affiliation{Laboratoire d'Information Quantique, Université libre de Bruxelles (ULB), Belgium}
	\author{Armin Tavakoli}
	\affiliation{Department of Physics and NanoLund, Lund University, Box 118, 22100 Lund, Sweden}
	
	\date{}
	
\begin{abstract}
The quantum prepare-and-measure scenario has been studied under various physical assumptions on the emitted states. Here, we first discuss how different assumptions are conceptually and formally related. We then identify one that can serve as a relaxation of all others, corresponding to a limitation on the one-shot accessible information of the state ensemble. This motivates us to study the optimal state discrimination probability of a source subject to these various physical assumptions. We derive general and tight bounds for states restricted by their quantum dimension, their vacuum component, an arbitrary uniform overlap, the magnitude of higher-dimensional signals and the experimenter's trust in their device. Our results constitute a first step towards a more unified picture of semi-device-independent quantum information processing. 
\end{abstract}
	
	\maketitle

	
\section{Introduction}

\begin{figure}[t!]
	\centering
	\includegraphics[width=0.8\columnwidth]{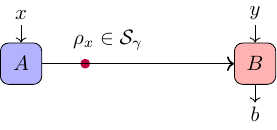}
	\caption{\emph{Prepare-and-measure scenario.} Different SDI assumptions, $\gamma$, specify a restriction on the states that Alice can send to Bob. The set of allowed states under a given SDI assumption is denoted $\mathcal{S}_\gamma$.}\label{FigScenario}
\end{figure}

The prepare-and-measure (PM) scenario, Fig.~\ref{FigScenario}, formalizes the simplest instance of a quantum communication experiment. A sender, Alice, encodes classical data into quantum systems which are sent to a receiver, Bob, who performs measurements to extract information. Protocols like BB84 and its many descendants \cite{Gisin2002} are archetypal examples of PM protocols, featuring fully characterized preparation and measurement devices. The advent of quantum technologies, and cryptographic applications in particular, has motivated the study of PM scenarios where devices are left uncharacterized up to some natural physical assumption on the preparation device. This is often referred to as semi-device-independent (SDI) quantum information.  

The most common SDI assumption restricts the Hilbert space dimension of the states. Quantum systems can create correlations that cannot be simulated by classical systems of the same dimension \cite{Wehner2008,Gallego2010,Brunner2013, Tavakoli2015}. This quantum-classical separation enables SDI quantum information protocols for quantum key distribution \cite{Pawlowski2011, Woodhead2015}, quantum random number generation \cite{Li2012, Lunghi2015}, self-testing and certification \cite{Tavakoli2018, Farkas2019, Tavakoli2020,Navascues2023,Egelhaaf2024} and entanglement detection \cite{Abbott2018, Moreno2021, bakhshinezhad2024}.  Dimension-restricted communication has also been studied when the parties can additionally share unbounded entanglement \cite{Pauwels2021, Pauwels2022, Vieira2023}.

The dimension represents the number of relevant degrees of freedom under the control of the experimenter. However, this is neither observable nor easy to precisely characterize. These shortcomings, which are especially salient for cryptographic applications, have partly motivated alternative communication assumptions. For example, the ``almost dimension'' approach assumes the states nearly, but not exactly, admit a $d$-dimensional representation \cite{AlmostQudit}.  Other proposals move away from the dimension entirely. An approach particularly well-suited to optical platforms supposes a limit on the photon excitations of the states, measured through the magnitude of the non-vacuum component  \cite{VanHimbeeck2017}. This has e.g.~been used for random number generation, in theory \cite{VanHimbeeck2019, senno2021} and practice \cite{Rusca2019, Rusca2020, Tebyanian2021, Avesani2021}. Another approach is to bound the pairwise overlap between the states emitted by Alice \cite{Brask2017, Wang2019}, which also  has been used in various protocols \cite{Ioannou2022, Ioannou2022b, Carceller2022, Shi2019, Fan2022}. Yet another communication assumption is a bound on the fidelity with which Alice prepares the specific states that she ideally intends to send to Bob \cite{Distrust}.  This may be viewed as a relaxation of the perfect preparation assumption used in one-sided device-independent approaches \cite{Ivan2017}. For brevity, we shall refer to these different assumptions as (i) the dimension restriction, (ii) entanglement-assisted dimension restriction,  (iii) the vacuum component restriction, (iv) the overlap restriction,  (v) almost dimension restriction, and (vi) the distrust restriction. 
Approaches (iii)-(vi) are primarily motivated by practical considerations, and (i)-(ii) partly also by the fundamental interest in comparing classical and quantum systems.

In contrast to the above assumptions that relate directly to the physical or quantum aspects of the preparation, a communication framework was introduced in \cite{InfoPaper1, InfoPaper2,Chaturvedi2020}, in which the only assumption has an information-theoretic interpretation. Roughly speaking, it quantifies how much knowledge could in principle be gained about Alice's input by measuring her states. This information restriction on Alice's preparations can, in general, neither be directly deduced from the setup nor accurately bounded by measuring a suitable observable. Operationally, it can be interpreted as the best quantum state discrimination possible on Alice's states \cite{Konig2009}. In this way, it provides an avenue to quantify the information cost of creating correlations between Alice and Bob.  

In this paper, we begin by structuring this landscape of quantum communication assumptions and then identifying connections between the different frameworks, but argue that they admit no generic hierarchy of relations in terms of the supported quantum correlations. However, we observe that all assumptions admit a one-way connection to information.  To make this connection explicit, we address the independently interesting question of determining the information capacity of quantum communication subjected to any of the considered restrictions. This amounts to bounding the best success probability in quantum state discrimination compatible with states limited by the various communication assumptions. For all assumptions, this is achieved analytically, in complete generality  and, in most cases,  provably tightly.

\section{Overview of common SDI assumptions}
 Consider the prepare-and-measure scenario in Fig.~\ref{FigScenario}. Alice privately selects an input $x$ and encodes it in a quantum message $\rho_x$, sent to Bob over a noiseless channel, who performs a decoding measurement $\{M_{b|y}\}_{y}$ depending on $y$, yielding an outcome $b$. The resulting correlations are given by Born's rule, $p(b|x,y)=\tr\left(\rho_x M_{b|y}\right)$. Naturally, if no restriction is imposed on the systems $\rho_x$, Alice may simply send $x$ over the channel, and Bob can simulate any $p(b|x,y)$. An assumed restriction on the devices is thus needed to limit the set of achievable correlations.
Many assumptions, usually on the preparation device or, equivalently, on the channel connecting Alice and Bob have been studied in the literature. We summarize the most common ones below.

\textbf{(i) Dimension \cite{Wehner2008,Gallego2010}. }
The state $\rho_x$ is assumed to live in a  Hilbert of fixed dimension $d$, that is, $\rho_x\in\mathcal{L}(\mathbb{C}^d)$, where $\mathcal{L}(\mathbb{C}^d)$ denotes the set of linear operators on $\mathbb{C}^d$. W.l.g.~one can also limit the measurement to be $d$-dimensional, but one can notably not assume them projective \cite{Smania2020, Martinez2023, Feng2023}. 

\textbf{(ii) Entanglement-assisted dimension \cite{Pauwels2021,Pauwels2022,Vieira2023}.} Alice and Bob share an entangled state $\phi_{AB}$. Alice encodes $x$ in her part of the entangled state using a quantum channel $\Lambda_x$ with fixed output dimension $d$ (the dimension of her message), and sends the output to Bob. Bob measures the total state $\rho_x=(\Lambda_x\otimes \openone)[\phi_{AB}]$. One can generally not restrict the local dimension of $\phi_{AB}$ to message dimension $d$ \cite{Guo2023} and no upper bound on it is known.

\textbf{(iii) Vacuum component \cite{VanHimbeeck2017,VanHimbeeck2019}.} Define the the Hamiltonian $H=\openone-\ketbra{0}$ and associate the state $\ket{0}$ to vacuum (no photon). A restriction, $\omega$, is assumed on photon exctitation of Alice's states as $\tr(H\rho_x) \leq \omega$.

\textbf{(iv) Overlaps \cite{Brask2017,Wang2019}.} The states $\rho_x$ are assumed to have purifications, $\ket{\psi_x}$, whose pairwise overlaps are bounded as $\left|\braket{\psi_x}{\psi_{x'}}\right|\geq a_{xx'}$ for some $a_{xx'}\in\mathbb{R}$.

\textbf{(v) Almost-dimension \cite{AlmostQudit}.}  Assume that there exists a $d$-dimensional space, with projector $\Pi_d$, in which nearly all of the support of Alice's  states is contained,
\begin{equation}\label{almostqudit}
	\tr(\rho_x \Pi_d)\geq 1-\eps~,
\end{equation}
for some small dimension-deviation parameters $\eps\geq 0$. The part of the state that is not supported on $\Pi_d$ corresponds to a deviation from the ideal $d$-dimensional system.

\textbf{(vi) Distrust \cite{Distrust}.}  Alice aims to prepare a state $\ket{\psi_x}$ but her preparation device realizes the lab state $\rho_x$. Her distrust in the device is limited through the fidelity between the target state and the lab state, $\bracket{\psi_x}{\rho_x}{\psi_x}\geq 1-\epsilon$, where $\epsilon\geq0$ is the distrust parameter. The lab states need not be of the same dimension as the target states.

\textbf{(*) Information \cite{InfoPaper1,InfoPaper2,Chaturvedi2020}.} Given the classical-quantum state $\rho_{XB}=\frac{1}{n}\sum_{x=1}^n \ketbra{x}\otimes \rho_x$, the conditional min-entropy is $H_\text{min}(X|B) = - \log_2(P_g)$ with
\begin{equation}\label{Pg}
	P_g\left(\{\rho_x\}\right)\equiv \max_{\{N_x\}} \frac{1}{n} \sum_{x=1}^n \tr\left(\rho_x N_x\right),
\end{equation}
where $n$ is the number of states and $\{N_x\}_x$ is a measurement. Thus, $P_g$ is the optimal probability of correctly guessing the classical value $x$ given the quantum state $\rho_x$ \cite{Konig2009}. 
The accessible information, measuring how much information the states $\rho_x$ convey about $x$, is then defined as the entropy difference $
	\mathcal{I}\equiv H_\text{min}(X)-H_\text{min}(X|B)=\log(n)+\log(P_g)$, where we assumed that $X$ is uniform.
The information restriction assumption introduced in \cite{InfoPaper1, InfoPaper2,Chaturvedi2020} is then the limit $\mathcal{I}\leq \alpha$ for some $\alpha\geq 0$, or equivalently, a limitation  $P_g\leq \frac{2^\alpha}{n}$.

\section{Role of shared randomness}
 In general, a communication assumption can be written as $\{\rho_x\}_x\in \mathcal{S}_\gamma$, for some selected set $\mathcal{S}_\gamma$, where $\gamma$ indexes the specific assumption parameters. Specifically, $\gamma=d$ for the (entanglement-assisted) dimension, $\gamma = \omega$ for the vacuum assumption, $\gamma= (a_{xx'})_{xx'}$ for the overlap assumption, $\gamma=(d,\epsilon)$ for the almost dimension, $\gamma = (\eps,\psi_x)_x$ for the distrust and $\gamma = \alpha$ for the information. The corresponding PM scenario can always be extended by allowing for shared randomness (SR) between Alice and Bob, leading to the correlations $p(b|x,y)= \sum_\lambda q_\lambda \tr\left(\rho_x^\lambda M_{b|y}^\lambda\right)$, where $\lambda$ denotes the shared randomness and $q$ is a distribution. Now, the assumption can be formulated in two different ways; either as \textit{peak-$\gamma$} or  \textit{average-$\gamma$}. Average-$\gamma$ means that the parameter assumption, $\gamma$, holds when averaging out $\lambda$,
\begin{equation}\label{eq:maxaverage}
\{\rho_x^\lambda\}_x\in \mathcal{S}_{\gamma_\lambda}\quad\text{ with  }\sum_\lambda q_\lambda \gamma_\lambda = \gamma.
\end{equation}
For instance, in the vacuum component framework, the states can sometimes have larger or smaller vacuum components $\omega^\lambda = \tr (\rho^\lambda_x H)$ but on average it respects the vacuum limit $\sum_\lambda q_\lambda \omega^\lambda \leq \omega$. Peak-$\gamma$ means that the assumption holds also when conditioning on $\lambda$, 
\begin{equation}\label{eq:maxpeak}
	\{\rho_x^\lambda\}_x\in \mathcal{S}_\gamma\,,\quad \forall \lambda\,.
\end{equation}
Continuing the example, the states may be different for each $\lambda$ but their vacuum component always respects the limit $\tr(\rho_x^\lambda H) \leq \omega$ for all $\lambda$.

The {peak-$\gamma$} assumption is natural for all cases considered above. At the level of the correlations, it corresponds simply to taking the convex-hull of the set of correlations without SR. The {average-$\gamma$} restriction has been studied explicitly for the vacuum \cite{VanHimbeeck2017}, almost qudit \cite{AlmostQudit} and distrust \cite{Distrust} assumptions. It has also been studied for Bell scenarios with dimension assumptions \cite{Gribling2018}; see also Appendix C in Ref.~\cite{AlmostQudit}.

It is also relevant to distinguish between whether the states $\rho_x^\lambda$ are assumed pure or mixed. Depending on the assumption, this can change the set of correlations \cite{InfoPaper2, Abbott2021}. While purity can be assumed w.l.g for (entanglement-assisted) dimension (by exploiting shared randomness, every mixed state admits a decomposition into pure states, each satisfying the assumption), there is neither a proof nor a counterexample of the same holding for the other assumptions. 

For all assumptions, the set of correlations without SR under assumption $\gamma$ is strictly contained in the set with peak-$\gamma$ SR, which itself is strictly contained in the set for average-$\gamma$ SR. 	In constrast, for the information assumption, the correlations without SR, with peak-$\gamma$ SR, and with average-$\gamma$ SR are all equivalent. Following eqs.~\eqref{Pg}, \eqref{eq:maxaverage} and \eqref{eq:maxpeak}, these three sets are defined as those corresponding to a source sending, respectively, states $\rho_x$ satisfying $P_g(\{\rho_x\})\leq \gamma$, states $\rho_x^\lambda$ satisfying $P_g(\{\rho_x^\lambda\})\leq \gamma$ for all $\lambda$, and states $\rho_x^\lambda$ satisfying $\sum_\lambda q(\lambda) P_g(\{\rho_x^\lambda\})\leq \gamma$. But the SR can always be incorporated in the emitted states themselves without increasing their information content \cite{InfoPaper2}. Indeed, simply define as emitted states the cq-states $\tilde \rho_x = \sum_\lambda q(\lambda) |\lambda\rangle\langle \lambda|\otimes \rho_x^\lambda$, whose guessing probability is $P_g(\{\tilde \rho_x\}) = \sum_\lambda q(\lambda) P_g(\{\rho_x^\lambda\})\leq \gamma$. Any peak-$\gamma$ or average-$\gamma$ strategy can thus be recast as an equivalent strategy that does not feature SR and which satifies the information restriction $P_g\leq \gamma$.

Note that starting from the information restriction $\mathcal{I} = \log(n)+\log(P_g(\{\rho_x\}))\leq \alpha\equiv \gamma$, rather than directly from the guessing probability, there are two ways to extend it from the no SR case to the average-$\gamma$ SR. One possibility is to assume for each $\lambda$ that $\mathcal{I}_\lambda \leq \alpha_\lambda$ with $\sum_\lambda q(\lambda) \alpha_\lambda = \alpha$, i.e., the average is taken at the level of the information quantity itself. The other possibility is to take the average at the level of the guessing probability and define the information bound as a bound $\log(n)+\log(\sum_\lambda q(\lambda) P_g(\{\rho_x^\lambda\}))\leq \alpha$, which is equivalent to the average guessing probability bound $\sum_\lambda q(\lambda) P_g(\{\rho_x^\lambda\})\leq 2^\alpha/n $. This corresponds to the situation described above and to the choice made in \cite{InfoPaper1, InfoPaper2}.
	
\section{Connecting the assumptions}
\begin{figure}
	\centering
	\includegraphics[width=\columnwidth]{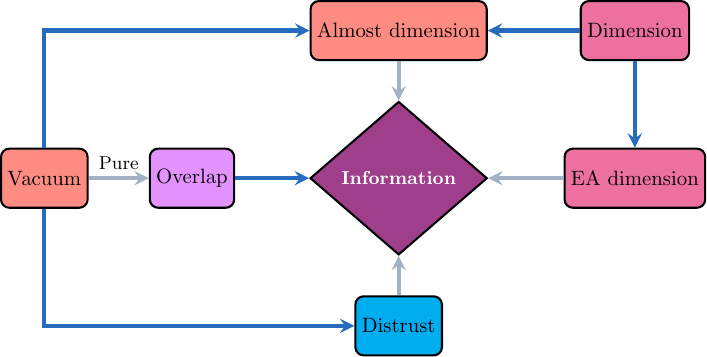}
	\caption{Relation between various SDI assumptions. Blue (grey) arrows indicate that one assumption is a special case (relaxation) of the other. The overlap relaxation of the vacuum only holds for pure states. }\label{FigRelations}
\end{figure}

 In Fig.~\ref{FigRelations}, we summarize the relations between the various assumptions, as discussed below. 
 \paragraph{Dimension-based assumptions.}
 The dimension is a special case of both the entanglement-assisted dimension and the almost dimension. In the former, we need only to restrict to sharing separable states, while in the latter we  just set the dimensional deviation  in Eq.~\eqref{almostqudit} to $\eps=0$. However,  these two are independent, and therefore incomparable, generalizations of the dimension restriction. 
\paragraph{Vacuum restriction.}
The vacuum component restriction can be seen as a limiting case of an almost dimension restriction. The latter is motivated as a correction to exact dimension restrictions, which are meaningful only when $d\geq 2$. However, an almost dimension restriction can in principle also be defined for $d=1$. The projector $\Pi_d$ is then a pure state, which we call the vacuum $\ket{0}$. This reduces Eq.~\eqref{almostqudit} to the vaccum component assumption. Consequently, the methodology developed in Ref.~\cite{AlmostQudit} for bounding correlations under almost $d$-dimensional systems can also be applied to analyze correlations under a vacuum component restriction \cite{Energy}. This is also a useful observation because, in the absence of such methods, the SDI protocols based on vacuum component restrictions have so far been limited to using just two states (analytically solvable) \cite{VanHimbeeck2017, VanHimbeeck2019}. 
Interestingly, the vacuum component restriction can equally be viewed as a limiting case of the distrust restriction. Indeed, the experimenter selects all target states to be vacuum, independently of $x$, i.e.~$\ket{\psi_x}=\ket{0}$.  This means that the numerical methods for distrust restricted correlations, developed in \cite{Distrust}, also can be used to analyze the case of restricted vacuum components. 

\paragraph{Almost dimension and distrust}
Does an almost dimension restriction admit any meaningful connection to a distrust restriction? The almost dimension too can be reformulated as a fidelity condition. The definition of fidelity for mixed states is $F(\sigma,\tau)=\left(\tr\sqrt{\sqrt{\sigma}\tau\sqrt{\sigma}}\right)^2$. For arbitrary positive operators $A$ and $B$, it follows that $F(A,BAB)=\tr(AB)^2$. The almost dimension assumption can then be interpreted as the fidelity between $\rho_x$ and its normalized projection onto the $d$-dimensional subspace $\Pi_d$. That is, $F\left(\rho_x,\frac{\Pi_d\rho_x\Pi_d}{\tr(\rho_x \Pi_d)}\right)=\tr(\rho_x \Pi_d)\geq 1-\eps$ via Eq.~\eqref{almostqudit}.  What decisively distinguishes it from the distrust assumption is that $\Pi_d$ cannot depend on $x$, as this would defeat the notion of the ensemble $\{\rho_x\}$ approximating a $d$-dimensional system.
\paragraph{Overlaps.}
Finally, the overlap restriction can be viewed as a relaxation of a vacuum restriction in the special case where all states $\ket{\psi_x}$ are pure. The vacuum component restriction $\tr(\ketbra{\psi_x}(\openone - \ketbra{0}))\leq \omega$ imposes a constraint on minimal overlap of the lab states, $\abs{\braket{\psi_x}{0}}\geq \sqrt{1-\omega}$. This restriction, on the minimal overlap between the lab states and the fixed reference state $\ket{0}$, in turn, implies a restriction on the minimal pairwise overlaps between the lab states. We will soon return to quantifying this.

\paragraph{Information.}

While the above assumptions in general are not comparable, they can all be linked to the notion of restricted accessible information (see Fig.~\ref{FigRelations}). This is possible because the information restriction has no privileged state spaces but only concerns how operationally useful the states are for carrying the information.  Specifically, there exist fundamental limits on the amount of information an $n$-state quantum ensemble can carry when restricted by any one of the assumptions under pure states. This leads directly to a bound under average-$\gamma$ SR in all cases except  entanglement-assisted dimension, for which the bound necessarily diverges unless one restricts to peak-$\gamma$ SR.

\section{Information cost}
Determining the largest accessible information that can be carried by an ensemble of $n$ pure states limited by any one of the trust assumptions amounts to evaluating an upper limitation on the ability to use such states for state discrimination, see Eq.~\eqref{Pg}. We express this as
\begin{equation}
P_g^*\equiv \max_{\{\psi_x\}\in \mathcal{S}_\gamma} P_g(\{\psi_x\}),
\end{equation}
where $\psi_x=\ketbra{\psi_x}$ is pure. We now address this problem for all considered settings. 
Then, we show how the results are valid also under SR.

\textbf{Dimension.} For completeness, we first rederive the known fact that any dimension-restricted ensemble, i.e.~$\mathcal{S}=\mathcal{L}(\mathbb{C}^d)$, carries at most $\log d$ bits \cite{Nayak1999}. Taking $n\geq d$ and using the fact that for any measurement $N_x$ the corresponding optimal states $\psi_x$ are projectors on the eigenvector of $N_x$ with the largest eigenvalue, we obtain  $P_g^*= \max_{\{N_x\}}\frac{1}{n}\sum_x \lambda_\text{max}(N_x) \leq \max_{\{N_x\}}\frac{1}{n}\sum_x\tr(N_x)=\frac{d}{n}$, where we have used that $\sum_x N_x=\openone_d$. The information cost becomes $\mathcal{I}\leq \log(d)$, independently of $n$. This bound can be saturated trivially by $d$ states forming a basis of $\mathbb{C}^d$. We note that this also follows directly from the fact that the signaling dimension \cite{DallArno2017} of quantum and classical systems of dimension $d$ are equal \cite{Frenkel2015}.

\textbf{Entanglement-assisted dimension.} The set of states $\psi_x$ generated via entanglement-assisted quantum communication obeys no-signaling, meaning that Bob's local state is independent of Alice's operation, i.e.~$\psi^B\equiv \tr_A(\psi_x)$ for all $x$. We can thus restrict to bipartite states with a constant marginal on  Bob. Then, for $n\geq d^2$ \cite{Pauwels2021}, $P_g^*\leq \max \frac{d}{n}\sum_x \tr(\psi_x^B N^B_x)=\max \frac{d}{n}\sum_x \tr(\psi_B N^B_x)=\frac{d^2}{n}$. Here, we used that the Schmidt number of $\psi_x$ is at most $d$ which implies $\tr(\psi_x N_x)\leq d \tr(\psi_x^B N_x^B)$, where $N_x^B=\tr_A(N_x)$. In the last step, we used that $\sum_x N_x^B=\tr_A \sum_x N_x=\tr_A(\openone_d\otimes \openone)=d \openone$.  The associated information cost is $\mathcal{I}\leq 2\log(d)$. This can be saturated using a $d$-dimensional dense coding protocol \cite{Bennett1992}.

\textbf{Overlaps.} When $n=2$, the overlap of two pure states is one-to-one with their accessible information. This follows from the derivation of the Helstom bound for two-state discrimination \cite{Helstrom1969}. In this sense, the overlap is a special case of the information restriction (thus the connection in Fig.~\ref{FigRelations}). For $n>2$ and a uniform overlap constraint $\braket{\psi_x}{\psi_{x'}}\geq a$, we can use the observation that for ensembles that are equiprobable and equiangular \cite{Krovi2015}  the optimal measurement for state discrimination is the so-called pretty good measurement \cite{Montanaro2007}, defined as $N_x=S^{-1/2}\psi_xS^{-1/2}$ where $S=\sum_x \psi_x$. This leads to
\begin{equation}
P_g \leq \frac{1}{n^2}\left((n-1)\sqrt{1-a}+\sqrt{(n-1)a+1}\right)^2 \,.
\end{equation} 

\textbf{Vacuum component.}  W.l.g we can select the amplitude associated with the vacuum $|0\rangle$ to be real and then use the simulation technique of \cite{Mckague2009} to embed the non-vacuum components in a sufficiently high-dimensional real-valued Hilbert space. The reachable state space therefore forms a cone around the vacuum state with radius $\sqrt{1-\omega}$. We optimally choose all states on the boundary of this cone, i.e.~$\sqrt{1-\omega}=\braket{0}{\psi_x}$. The value of $P_g$ is invariant under permutations of the label $x$. From any optimal solution, we can then always form a solution where all the states have identical overlaps $\braket{\psi_x}{\psi_{x'}}=a$ for all $x\neq x'$. Indeed, consider the direct sum states $|\Psi_x\rangle = \frac{1}{\sqrt{N}}\oplus_{k} \sigma_k |\psi_x\rangle$ where the sum $k$ runs over the $N$ permutations $\sigma_k$ of the input labels $x$. Then the states $|\Psi_x\rangle$ have now equal overlaps, guessing probability $P_g$ at least equal to the original one, and all have an overlap $\sqrt{1-\omega}$ with the vacuum state $\frac{1}{\sqrt{N}} \oplus_k |0\rangle$ (which can be unitarily mapped to the original vacuum state $|0\rangle\oplus 0 \ldots \oplus 0$ if desired).  

Given our result for the overlap assumption above, to evaluate $P_g$, we only need to minimize the overlap $a$. The Gram matrix associated with the states $\{\ket{\psi_1},\ldots,\ket{\psi_{n}},\ket{0}\}$ is $G=\begin{pmatrix}
A & B\\B^T & 1
\end{pmatrix}$, where $A$ is $n\times n$ with $1$ on the diagonals and $a$ on the off-diagonals, and $B$ is $n\times 1$ with all entries $\sqrt{1-\omega}$. 

It is positive semidefinite by construction. We therefore compute $a^*=\min_{G\succeq 0} a$. Using Schur complements, the eigenvalues of $G$ can be evaluated analytically. From the smallest one, we obtain the result $a^*=1-\frac{n}{n-1}\omega$. Combining this with the pretty good measurement leads to the bound
\begin{equation}\label{vacuumbound}
P_g \leq \frac{1}{n}\left(\sqrt{\omega(n-1)}+\sqrt{1-\omega}\right)^2,
\end{equation}
valid when $0\leq \omega\leq \frac{n-1}{n}$, while $P_g=1$ otherwise.  It can be saturated by construction. The corresponding accessible information increases monotonically in $n$. The reason is that even if $\omega$ is very small, since there is no dimension restriction, we can always choose the small non-vacuum component of each state orthogonal to that of all the other states, thus increasing the information.

\textbf{Almost dimension and distrust.} For bounding the information capacity associated with almost dimension-restricted or distrust-restricted states, we rely on a useful operator inequality. Specifically, we observe that the method in Ref.~\cite{Tavakoli2024} for proving Lemma 1 can be recycled to prove the following more general statement. Let $\ket{\phi}$ be any state such that $\bracket{\phi}{\Pi_d}{\phi}\geq 1-\eps$. Then, $\ketbra{\phi} \preceq (1+\mu)\tilde{\sigma}+h(\eps,\mu)\openone_D$ for every $\mu\geq -1$, where $\tilde{\sigma}=\frac{\Pi_d \ketbra{\phi}  \Pi_d}{\bracket{\phi}{\Pi_d}{\phi}}$ and $h(\eps,\mu)=(\sqrt{\mu^2+4\varepsilon(1+\mu)}-\mu)/2$.  We apply this independently to each pure $\psi_x$ in Eq.~\eqref{Pg} and optimally choose each associated $\mu$ to be identical. Using that $n$ states span an $n$-dimensional subspace, we can restrict to $n$-dimensional POVMs and obtain $
P_g^*\leq (1+\mu) \left[\max_{\{N_x\}}\frac{1}{n}\sum_x \tr(\tilde{\psi}_x N_x)\right] +h(\mu,\eps)$. In the case of the almost dimension assumption, $\tilde{\psi}_x$ are $d$-dimensional and hence the maximization is bounded by  $d/n$. In the case of distrust assumption, $\Pi_d$ is simply replaced by $\ketbra{\psi_x}$ and we have $\tilde{\psi}_x=\ketbra{\psi_x}$. Therefore, the maximisation becomes just $P_g(\{\psi_x\})$ for the target ensemble. Let $P_g^0$ denote the value of the maximization for either the almost dimension case or the distrust case. Minimising the RHS over $\mu$, we find 
\begin{equation} \label{eq:info_fromLemma}
P_g^\varepsilon \leq P_g^0+(1-2P_g^0)\varepsilon+2\sqrt{P_g^0(1-P_g^0)}\sqrt{\varepsilon(1-\varepsilon)}.
\end{equation}
Note that in the special case of $d=1$, for the almost dimension, for which $P_g^0=\frac{1}{n}$, this bound reduces to that obtained for vacuum component restrictions in Eq.~\eqref{vacuumbound}. We have not been able to prove that the bound is tight in general, but for every numerical case study conducted, we find that the bound is indeed tight for the almost dimension assumption \cite{MATLAB}. For the distrust assumption, the bound is generally not tight unless the target states are optimal for state discrimination \cite{MATLAB}.

It can be straightforwardly shown that the above information capacity bounds hold also when SR is included. First, from the linearity of $P_g$, our results also hold for the peak-$\gamma$ SR assumption. Second, we note that the bound on $P_g$ for every assumption except entanglement-assisted dimension is concave in each of the respective assumption parameters (the dimension $d$, the energy $\omega$, the overlap $a$, distrust parameter $\epsilon$, almost qudits $(\epsilon,d)$\footnote{It is straightforward to verify the concavity of the bivariate function \eqref{eq:info_fromLemma} from the negativity of the Hessian.}). 
This implies that our results also hold for the average-$\gamma$ assumption, for all but the entanglement-assisted dimension\footnote{For example, setting $N=30$, with an entanglement-assisted qutrit, a dense coding strategy achieves $P_g=9/30$. However, one can send an \emph{average} qutrit by mixing a qubit with probability $2/3$ and a 5-dimensional system with probability $1/3$. Using dense coding strategies, one can achieve a guessing probability of $P_g =  11/30$.}.

An independently interesting consequence of this result is to bound the information capacity of $n$ optical coherent states with a limited average photon number but arbitrary phase. Recall the coherent state $\ket{\alpha}=e^{-\frac{|\alpha|}{2}}\sum_{k=0}^\infty \frac{|\alpha|^ke^{i\theta k }}{\sqrt{k!}}\ket{k}$. For  small average photon numbers, $N=|\alpha|^2$, this can be seen as an almost qubit with $\eps=1-e^{-|\alpha|}(1+|\alpha|^2)\approx \sqrt{N}$. Inserting this in \eqref{eq:info_fromLemma} and then computing $\mathcal{I}$ yields the desired bound.

\section{Final remarks}
Quantum communication assumptions in the prepare-and-measure scenario can be divided into two classes. The first class limits the weight of the state on various subspaces. These can correspond to the available degrees of freedom (dimension, almost dimension and entanglement-assisted dimension), the vacuum subspace, the subspace spanned by each of the other states of Alice (overlap assumption), the subspace corresponding to her target states (distrust assumption) etc. The second class does not favor any particular subspaces but is instead concerned with limiting the capacity of the states w.r.t.~a specific operational task. This task can for instance be state discrimination, as in the considered information restriction setting, but can in principle be arbitrary. Developing a formalism and methodology for correlations obtained under assumptions on bounded subspace weights and  capacity restrictions are natural next steps towards a unified picture of SDI. This is important because although we have established several tight information capacity relations, this does not imply that the set of correlations under a given communication assumption can w.l.g.~be substituted with that obtained from the associated informationally restricted communication. In this context, we also note that while the set of correlations without SR, with average-$\gamma$ SR, and with peak-$\gamma$ SR can be distinct depending on the assumption, they all relax to the same set of information-constrained correlations if in the latter the SR is taken at the level of $P_g$, i.e., $P_g = \sum_\lambda q_\lambda P_g (\{\rho_x^\lambda\}_x) \leq \gamma$, as considered here and in \cite{InfoPaper1,InfoPaper2}. This would not be the case for an alternative definition, where the information itself is averaged over the SR, i.e., $I = \log(n) + \sum_\lambda q_\lambda \log P_g(\{\rho_x^\lambda\}_x) \leq \gamma$. Developing a formalism for the latter case is an interesting open question.

\begin{acknowledgments}	
A.T. is supported by the Wenner-Gren Foundation, by the Knut and Alice Wallenberg Foundation through the Wallenberg Center for Quantum Technology (WACQT) and the Swedish Research Council under Contract No.~2023-03498. J.P. acknowledges support from the Swiss National Science Foundation via the NCCR-SwissMap. S.P. acknowledges funding from the VERIqTAS project within the QuantERA II Programme that has received funding from the European Union's Horizon 2020 research and innovation program under Grant Agreement No 101017733 and the F.R.S-FNRS Pint-Multi program under Grant Agreement R.8014.21, from the European Union's Horizon Europe research and innovation program under the project ``Quantum Security Networks Partnership'' (QSNP, grant agreement No 101114043),  from the F.R.S-FNRS through the PDR T.0171.22, from the FWO and F.R.S.-FNRS under the Excellence of Science (EOS) program project 40007526, from the FWO through the BeQuNet SBO project S008323N, from the Belgian Federal Science Policy through the contract RT/22/BE-QCI and the EU ``BE-QCI'' program.
S.P. is a Research Director of the Fonds de la Recherche Scientifique -- FNRS.
\end{acknowledgments}

\bibliographystyle{quantum}

\bibliography{references_infocost}

\begin{thebibliography}{10}

\bibitem{Gisin2002}
Nicolas Gisin, Gr\'egoire Ribordy, Wolfgang Tittel, and Hugo Zbinden.
\newblock ``Quantum cryptography''.
\newblock \href{https://dx.doi.org/10.1103/RevModPhys.74.145}{Rev. Mod. Phys. {\bf 74}, 145--195}~(2002).

\bibitem{Wehner2008}
Stephanie Wehner, Matthias Christandl, and Andrew~C Doherty.
\newblock ``Lower bound on the dimension of a quantum system given measured data''.
\newblock \href{https://dx.doi.org/10.1103/PhysRevA.78.062112}{\pra {\bf 78}, 062112}~(2008).

\bibitem{Gallego2010}
Rodrigo Gallego, Nicolas Brunner, Christopher Hadley, and Antonio Ac\'{\i}n.
\newblock ``Device-independent tests of classical and quantum dimensions''.
\newblock \href{https://dx.doi.org/10.1103/PhysRevLett.105.230501}{Phys. Rev. Lett. {\bf 105}, 230501}~(2010).

\bibitem{Brunner2013}
Nicolas Brunner, Miguel Navascu\'es, and Tam\'as V\'ertesi.
\newblock ``Dimension witnesses and quantum state discrimination''.
\newblock \href{https://dx.doi.org/10.1103/PhysRevLett.110.150501}{Phys. Rev. Lett. {\bf 110}, 150501}~(2013).

\bibitem{Tavakoli2015}
Armin Tavakoli, Alley Hameedi, Breno Marques, and Mohamed Bourennane.
\newblock ``Quantum random access codes using single $d$-level systems''.
\newblock \href{https://dx.doi.org/10.1103/PhysRevLett.114.170502}{Phys. Rev. Lett. {\bf 114}, 170502}~(2015).

\bibitem{Pawlowski2011}
Marcin Paw\l{}owski and Nicolas Brunner.
\newblock ``Semi-device-independent security of one-way quantum key distribution''.
\newblock \href{https://dx.doi.org/10.1103/PhysRevA.84.010302}{Phys. Rev. A {\bf 84}, 010302}~(2011).

\bibitem{Woodhead2015}
Erik Woodhead and Stefano Pironio.
\newblock ``Secrecy in prepare-and-measure clauser-horne-shimony-holt tests with a qubit bound''.
\newblock \href{https://dx.doi.org/10.1103/PhysRevLett.115.150501}{Phys. Rev. Lett. {\bf 115}, 150501}~(2015).

\bibitem{Li2012}
Hong-Wei Li, Marcin Paw\l{}owski, Zhen-Qiang Yin, Guang-Can Guo, and Zheng-Fu Han.
\newblock ``Semi-device-independent randomness certification using $n\ensuremath{\rightarrow}1$ quantum random access codes''.
\newblock \href{https://dx.doi.org/10.1103/PhysRevA.85.052308}{Phys. Rev. A {\bf 85}, 052308}~(2012).

\bibitem{Lunghi2015}
Tommaso Lunghi, Jonatan~Bohr Brask, Charles Ci~Wen Lim, Quentin Lavigne, Joseph Bowles, Anthony Martin, Hugo Zbinden, and Nicolas Brunner.
\newblock ``Self-testing quantum random number generator''.
\newblock \href{https://dx.doi.org/10.1103/PhysRevLett.114.150501}{Phys. Rev. Lett. {\bf 114}, 150501}~(2015).

\bibitem{Tavakoli2018}
Armin Tavakoli, Jędrzej Kaniewski, Tam\'as V\'ertesi, Denis Rosset, and Nicolas Brunner.
\newblock ``Self-testing quantum states and measurements in the prepare-and-measure scenario''.
\newblock \href{https://dx.doi.org/10.1103/PhysRevA.98.062307}{Phys. Rev. A {\bf 98}, 062307}~(2018).

\bibitem{Farkas2019}
M\'at\'e Farkas and Jędrzej Kaniewski.
\newblock ``Self-testing mutually unbiased bases in the prepare-and-measure scenario''.
\newblock \href{https://dx.doi.org/10.1103/PhysRevA.99.032316}{Phys. Rev. A {\bf 99}, 032316}~(2019).

\bibitem{Tavakoli2020}
Armin Tavakoli.
\newblock ``Semi-device-independent certification of independent quantum state and measurement devices''.
\newblock \href{https://dx.doi.org/10.1103/PhysRevLett.125.150503}{Phys. Rev. Lett. {\bf 125}, 150503}~(2020).

\bibitem{Navascues2023}
Miguel Navascu\'es, K\'aroly~F. P\'al, Tam\'as V\'ertesi, and Mateus Ara\'ujo.
\newblock ``Self-testing in prepare-and-measure scenarios and a robust version of wigner's theorem''.
\newblock \href{https://dx.doi.org/10.1103/PhysRevLett.131.250802}{Phys. Rev. Lett. {\bf 131}, 250802}~(2023).

\bibitem{Egelhaaf2024}
Sophie Egelhaaf, Jef Pauwels, Marco~Túlio Quintino, and Roope Uola.
\newblock ``Certifying measurement incompatibility in prepare-and-measure and bell scenarios''~(2024).
\newblock  \href{http://arxiv.org/abs/2407.06787}{arXiv:2407.06787}.

\bibitem{Abbott2018}
Armin Tavakoli, Alastair~A. Abbott, Marc-Olivier Renou, Nicolas Gisin, and Nicolas Brunner.
\newblock ``Semi-device-independent characterization of multipartite entanglement of states and measurements''.
\newblock \href{https://dx.doi.org/10.1103/PhysRevA.98.052333}{Phys. Rev. A {\bf 98}, 052333}~(2018).

\bibitem{Moreno2021}
George Moreno, Ranieri Nery, Carlos de~Gois, Rafael Rabelo, and Rafael Chaves.
\newblock ``Semi-device-independent certification of entanglement in superdense coding''.
\newblock \href{https://dx.doi.org/10.1103/PhysRevA.103.022426}{Phys. Rev. A {\bf 103}, 022426}~(2021).

\bibitem{bakhshinezhad2024}
Pharnam Bakhshinezhad, Mohammad Mehboudi, Carles~Roch i~Carceller, and Armin Tavakoli.
\newblock ``Scalable entanglement certification via quantum communication''~(2024).
\newblock  \href{http://arxiv.org/abs/2401.00796}{arXiv:2401.00796}.

\bibitem{Pauwels2021}
Armin Tavakoli, Jef Pauwels, Erik Woodhead, and Stefano Pironio.
\newblock ``Correlations in entanglement-assisted prepare-and-measure scenarios''.
\newblock \href{https://dx.doi.org/10.1103/PRXQuantum.2.040357}{PRX Quantum {\bf 2}, 040357}~(2021).

\bibitem{Pauwels2022}
Jef Pauwels, Armin Tavakoli, Erik Woodhead, and Stefano Pironio.
\newblock ``Entanglement in prepare-and-measure scenarios: many questions, a few answers''.
\newblock \href{https://dx.doi.org/10.1088/1367-2630/ac724a}{NJPq {\bf 24}, 063015}~(2022).

\bibitem{Vieira2023}
Carlos Vieira, Carlos de~Gois, Lucas Pollyceno, and Rafael Rabelo.
\newblock ``Interplays between classical and quantum entanglement-assisted communication scenarios''.
\newblock \href{https://dx.doi.org/10.1088/1367-2630/ad0526}{NJP {\bf 25}, 113004}~(2023).

\bibitem{AlmostQudit}
Jef Pauwels, Stefano Pironio, Erik Woodhead, and Armin Tavakoli.
\newblock ``Almost qudits in the prepare-and-measure scenario''.
\newblock \href{https://dx.doi.org/10.1103/PhysRevLett.129.250504}{Phys. Rev. Lett. {\bf 129}, 250504}~(2022).

\bibitem{VanHimbeeck2017}
Thomas Van~Himbeeck, Erik Woodhead, Nicolas~J. Cerf, Ra{\'{u}}l Garc{\'{i}}a-Patr{\'{o}}n, and Stefano Pironio.
\newblock ``Semi-device-independent framework based on natural physical assumptions''.
\newblock \href{https://dx.doi.org/10.22331/q-2017-11-18-33}{{Quantum} {\bf 1}, 33}~(2017).

\bibitem{VanHimbeeck2019}
Thomas~Van Himbeeck and Stefano Pironio.
\newblock ``Correlations and randomness generation based on energy constraints''~(2019).
\newblock  \href{http://arxiv.org/abs/1905.09117}{arXiv:1905.09117}.

\bibitem{senno2021}
Gabriel Senno and Antonio Acín.
\newblock ``Semi-device-independent full randomness amplification based on energy bounds''~(2021).
\newblock  \href{http://arxiv.org/abs/2108.09100}{arXiv:2108.09100}.

\bibitem{Rusca2019}
Davide Rusca, Thomas van Himbeeck, Anthony Martin, Jonatan~Bohr Brask, Weixu Shi, Stefano Pironio, Nicolas Brunner, and Hugo Zbinden.
\newblock ``Self-testing quantum random-number generator based on an energy bound''.
\newblock \href{https://dx.doi.org/10.1103/PhysRevA.100.062338}{Phys. Rev. A {\bf 100}, 062338}~(2019).

\bibitem{Rusca2020}
Davide Rusca, Hamid Tebyanian, Anthony Martin, and Hugo Zbinden.
\newblock ``Fast self-testing quantum random number generator based on homodyne detection''.
\newblock \href{https://dx.doi.org/10.1063/5.0011479}{Appl. Phys. Lett {\bf 116}, 264004}~(2020).

\bibitem{Tebyanian2021}
Hamid Tebyanian, Mujtaba Zahidy, Marco Avesani, Andrea Stanco, Paolo Villoresi, and Giuseppe Vallone.
\newblock ``Semi-device independent randomness generation based on quantum state's indistinguishability''.
\newblock \href{https://dx.doi.org/10.1088/2058-9565/ac2047}{QST {\bf 6}, 045026}~(2021).

\bibitem{Avesani2021}
Marco Avesani, Hamid Tebyanian, Paolo Villoresi, and Giuseppe Vallone.
\newblock ``Semi-device-independent heterodyne-based quantum random-number generator''.
\newblock \href{https://dx.doi.org/10.1103/PhysRevApplied.15.034034}{Phys. Rev. Appl. {\bf 15}, 034034}~(2021).

\bibitem{Brask2017}
Jonatan~Bohr Brask, Anthony Martin, William Esposito, Raphael Houlmann, Joseph Bowles, Hugo Zbinden, and Nicolas Brunner.
\newblock ``Megahertz-rate semi-device-independent quantum random number generators based on unambiguous state discrimination''.
\newblock \href{https://dx.doi.org/10.1103/PhysRevApplied.7.054018}{Phys. Rev. Appl. {\bf 7}, 054018}~(2017).

\bibitem{Wang2019}
Yukun Wang, Ignatius~William Primaatmaja, Emilien Lavie, Antonios Varvitsiotis, and Charles Ci~Wen Lim.
\newblock ``Characterising the correlations of prepare-and-measure quantum networks''.
\newblock \href{https://dx.doi.org/10.1038/s41534-019-0133-3}{npj Quantum Inf. {\bf 5}, 17}~(2019).

\bibitem{Ioannou2022}
Marie Ioannou, Maria~Ana Pereira, Davide Rusca, Fadri Gr{\"{u}}nenfelder, Alberto Boaron, Matthieu Perrenoud, Alastair~A. Abbott, Pavel Sekatski, Jean-Daniel Bancal, Nicolas Maring, Hugo Zbinden, and Nicolas Brunner.
\newblock ``Receiver-{D}evice-{I}ndependent {Q}uantum {K}ey {D}istribution''.
\newblock \href{https://dx.doi.org/10.22331/q-2022-05-24-718}{{Quantum} {\bf 6}, 718}~(2022).

\bibitem{Ioannou2022b}
Marie Ioannou, Pavel Sekatski, Alastair~A Abbott, Denis Rosset, Jean-Daniel Bancal, and Nicolas Brunner.
\newblock ``Receiver-device-independent quantum key distribution protocols''.
\newblock \href{https://dx.doi.org/10.1088/1367-2630/ac71bc}{NJP {\bf 24}, 063006}~(2022).

\bibitem{Carceller2022}
Carles Roch~i Carceller, Kieran Flatt, Hanwool Lee, Joonwoo Bae, and Jonatan~Bohr Brask.
\newblock ``Quantum vs noncontextual semi-device-independent randomness certification''.
\newblock \href{https://dx.doi.org/10.1103/PhysRevLett.129.050501}{Phys. Rev. Lett. {\bf 129}, 050501}~(2022).

\bibitem{Shi2019}
Weixu Shi, Yu~Cai, Jonatan~Bohr Brask, Hugo Zbinden, and Nicolas Brunner.
\newblock ``Semi-device-independent characterization of quantum measurements under a minimum overlap assumption''.
\newblock \href{https://dx.doi.org/10.1103/PhysRevA.100.042108}{Phys. Rev. A {\bf 100}, 042108}~(2019).

\bibitem{Fan2022}
Qin Fan, Meng-Yun Ma, Yong-Nan Sun, Qi-Ping Su, and Chui-Ping Yang.
\newblock ``Experimental certification of nonprojective quantum measurements under a minimum overlap assumption''.
\newblock \href{https://dx.doi.org/10.1364/OE.469225}{Opt. Express {\bf 30}, 34441--34452}~(2022).

\bibitem{Distrust}
Armin Tavakoli.
\newblock ``Semi-device-independent framework based on restricted distrust in prepare-and-measure experiments''.
\newblock \href{https://dx.doi.org/10.1103/PhysRevLett.126.210503}{Phys. Rev. Lett. {\bf 126}, 210503}~(2021).

\bibitem{Ivan2017}
Ivan \ifmmode \check{S}\else \v{S}\fi{}upi\ifmmode~\acute{c}\else \'{c}\fi{}, Paul Skrzypczyk, and Daniel Cavalcanti.
\newblock ``Measurement-device-independent entanglement and randomness estimation in quantum networks''.
\newblock \href{https://dx.doi.org/10.1103/PhysRevA.95.042340}{Phys. Rev. A {\bf 95}, 042340}~(2017).

\bibitem{InfoPaper1}
Armin Tavakoli, Emmanuel Zambrini~Cruzeiro, Jonatan Bohr~Brask, Nicolas Gisin, and Nicolas Brunner.
\newblock ``Informationally restricted quantum correlations''.
\newblock \href{https://dx.doi.org/10.22331/q-2020-09-24-332}{{Quantum} {\bf 4}, 332}~(2020).

\bibitem{InfoPaper2}
Armin Tavakoli, Emmanuel Zambrini~Cruzeiro, Erik Woodhead, and Stefano Pironio.
\newblock ``Informationally restricted correlations: a general framework for classical and quantum systems''.
\newblock \href{https://dx.doi.org/10.22331/q-2022-01-05-620}{{Quantum} {\bf 6}, 620}~(2022).

\bibitem{Chaturvedi2020}
Anubhav Chaturvedi and Debashis Saha.
\newblock ``Quantum prescriptions are more ontologically distinct than they are operationally distinguishable''.
\newblock \href{https://dx.doi.org/10.22331/q-2020-10-21-345}{Quantum {\bf 4}, 345}~(2020).

\bibitem{Konig2009}
Robert Konig, Renato Renner, and Christian Schaffner.
\newblock ``The operational meaning of min- and max-entropy''.
\newblock \href{https://dx.doi.org/10.1109/TIT.2009.2025545}{IEEE Trans. Inf. Theory {\bf 55}, 4337--4347}~(2009).

\bibitem{Smania2020}
Armin Tavakoli, Massimiliano Smania, Tamás Vértesi, Nicolas Brunner, and Mohamed Bourennane.
\newblock ``Self-testing nonprojective quantum measurements in prepare-and-measure experiments''.
\newblock \href{https://dx.doi.org/10.1126/sciadv.aaw6664}{Sci. Adv. {\bf 6}, eaaw6664}~(2020).

\bibitem{Martinez2023}
Daniel Mart{\'i}nez, Esteban~S. G{\'o}mez, Jaime Cari{\~{n}}e, Luciano Pereira, Aldo Delgado, Stephen~P. Walborn, Armin Tavakoli, and Gustavo Lima.
\newblock ``Certification of a non-projective qudit measurement using multiport beamsplitters''.
\newblock \href{https://dx.doi.org/10.1038/s41567-022-01845-z}{Nat. Phys. {\bf 19}, 190--195}~(2023).

\bibitem{Feng2023}
Lan-Tian Feng, Xiao-Min Hu, Ming Zhang, Yu-Jie Cheng, Chao Zhang, Yu~Guo, Yu-Yang Ding, Zhibo Hou, Fang-Wen Sun, Guang-Can Guo, Dao-Xin Dai, Armin Tavakoli, Xi-Feng Ren, and Bi-Heng Liu.
\newblock ``Higher-dimensional symmetric informationally complete measurement via programmable photonic integrated optics''~(2023).
\newblock  \href{http://arxiv.org/abs/2310.08838}{arXiv:2310.08838}.

\bibitem{Guo2023}
Yu~Guo, Hao Tang, Jef Pauwels, Emmanuel~Zambrini Cruzeiro, Xiao-Min Hu, Bi-Heng Liu, Yu-Feng Huang, Chuan-Feng Li, Guang-Can Guo, and Armin Tavakoli.
\newblock ``Experimental higher-dimensional entanglement advantage over qubit channel''~(2023).
\newblock  \href{http://arxiv.org/abs/2306.13495v2}{arXiv:2306.13495v2}.

\bibitem{Gribling2018}
Sander Gribling, David de~Laat, and Monique Laurent.
\newblock ``Bounds on entanglement dimensions and quantum graph parameters via noncommutative polynomial optimization''.
\newblock \href{https://dx.doi.org/10.1007/s10107-018-1287-z}{Math. Program. {\bf 170}, 5--42}~(2018).

\bibitem{Abbott2021}
Armin Tavakoli, Emmanuel~Zambrini Cruzeiro, Roope Uola, and Alastair~A. Abbott.
\newblock ``Bounding and simulating contextual correlations in quantum theory''.
\newblock \href{https://dx.doi.org/10.1103/PRXQuantum.2.020334}{PRX Quantum {\bf 2}, 020334}~(2021).

\bibitem{Energy}
Carles~Roch i~Carceller, Jef Pauwels, Stefano Pironio, and Armin Tavakoli.
\newblock ``Prepare-and-measure scenarios with photon-number constraints''~(2024).
\newblock  \href{http://arxiv.org/abs/2412.13000}{arXiv:2412.13000}.

\bibitem{Nayak1999}
A.~Nayak.
\newblock ``Optimal lower bounds for quantum automata and random access codes''.
\newblock In 40th Annual Symposium on Foundations of Computer Science (Cat. No.99CB37039).
\newblock \href{https://dx.doi.org/10.1109/SFFCS.1999.814608}{Pages 369--376}.
\newblock ~(1999).

\bibitem{DallArno2017}
Michele Dall'Arno, Sarah Brandsen, Alessandro Tosini, Francesco Buscemi, and Vlatko Vedral.
\newblock ``No-hypersignaling principle''.
\newblock \href{https://dx.doi.org/10.1103/physrevlett.119.020401}{Physical Review Letters {\bf 119}, 020401}~(2017).

\bibitem{Frenkel2015}
Péter~E. Frenkel and Mihály Weiner.
\newblock ``Classical information storage in an n-level quantum system''.
\newblock \href{https://dx.doi.org/10.1007/s00220-015-2463-0}{Communications in Mathematical Physics {\bf 340}, 563--574}~(2015).

\bibitem{Bennett1992}
Charles~H. Bennett and Stephen~J. Wiesner.
\newblock ``Communication via one- and two-particle operators on {Einstein-Podolsky-Rosen} states''.
\newblock \href{https://dx.doi.org/10.1103/PhysRevLett.69.2881}{Phys. Rev. Lett. {\bf 69}, 2881--2884}~(1992).

\bibitem{Helstrom1969}
Carl~W. Helstrom.
\newblock ``Quantum detection and estimation theory''.
\newblock \href{https://dx.doi.org/10.1007/BF01007479}{J. Stat. Phys. {\bf 1}, 231--252}~(1969).

\bibitem{Krovi2015}
Hari Krovi, Saikat Guha, Zachary Dutton, and Marcus~P. da~Silva.
\newblock ``Optimal measurements for symmetric quantum states with applications to optical communication''.
\newblock \href{https://dx.doi.org/10.1103/physreva.92.062333}{Phys. Rev. A {\bf 92}, 062333}~(2015).

\bibitem{Montanaro2007}
Ashley Montanaro.
\newblock ``On the distinguishability of random quantum states''.
\newblock \href{https://dx.doi.org/10.1007/s00220-007-0221-7}{Comm. Math. Phys. {\bf 273}, 619--636}~(2007).

\bibitem{Mckague2009}
Matthew McKague, Michele Mosca, and Nicolas Gisin.
\newblock ``Simulating quantum systems using real {Hilbert} spaces''.
\newblock \href{https://dx.doi.org/10.1103/PhysRevLett.102.020505}{Phys. Rev. Lett. {\bf 102}, 020505}~(2009).

\bibitem{Tavakoli2024}
Armin Tavakoli.
\newblock ``Quantum steering with imprecise measurements''.
\newblock \href{https://dx.doi.org/10.1103/physrevlett.132.070204}{Physical Review Letters {\bf 132}, 070204}~(2024).

\bibitem{MATLAB}
``The matlab script that was used to verify this can be found on \url{https://github.com/jefpauwels/SDISeesaw.}''.

\end{thebibliography}

\end{document}